# Electro-thermal equivalent 3D Finite Element model of a single neuron

Illaria Cinelli, Michel Destrade, Maeve Duffy, Peter McHugh
NUI Galway, University Road, Galway, Ireland

*Abstract— Objective:* We propose a novel approach for modelling the inter-dependence of electrical and mechanical phenomena in nervous cells, by using electro-thermal equivalences in finite element (FE) analysis so that existing thermo-mechanical tools can be applied. *Methods:* First, the equivalence between electrical and thermal properties of the nerve materials is established, and results of a pure heat conduction analysis performed in Abaqus CAE Software 6.13-3 are validated with analytical solutions for a range of steady and transient conditions. This validation includes the definition of equivalent active membrane properties that enable prediction of the action potential. Then, as a step towards fully coupled models, electro-mechanical coupling is implemented through the definition of equivalent piezoelectric properties of the nerve membrane using the thermal expansion coefficient, enabling prediction of the mechanical response of the nerve to the action potential. *Results:* Results of the coupled electro-mechanical model are validated with previously published experimental results of deformation for the squid giant axon, crab nerve fibre and garfish olfactory nerve fibre. *Conclusion:* A simplified coupled electro-mechanical modelling approach is established through an electro-thermal equivalent FE model of a nervous cell for biomedical applications. *Significance:* One of the key findings is the mechanical characterization of the neural activity in a coupled electro-mechanical domain, which provides insights into the electro-mechanical behaviour of nervous cells, such as thinning of the membrane. This is a first step towards modelling 3D electro-mechanical alteration induced by trauma at nerve bundle, tissue and organ levels.

*Keywords:* Coupled electro-mechanical modelling, finite element modelling, equivalences, Cable Equation, Heat Equation, nervous cell, membrane.

## I. INTRODUCTION

The complexity of numerical modelling of the biomechanical behaviour of biological tissue is constantly increasing due the use of more realistic geometries and the incorporation of the non-linear behaviour of the tissue at different levels (such as cells and fibres) [1], [2]. Moreover, the increase in degrees of freedom comes with a higher computational cost [1]–[3].

In contrast, numerical modelling of the electrical activity of biological tissue is based on completely different principles, and the requirements differ in terms of equations, boundary conditions and discretisation [2], [3]. Therefore, the electrical activity of tissue is generally modelled using dedicated electrical modelling software codes [2], [4].

Detailed electrophysiological models of cellular and tissue processes [3] are now being linked to a large body of existing research for modelling the spread of electrical excitation throughout biological tissue (such as nerve and cardiac tissue) [1], [2]. From the perspective of *combining* the electrical and biomechanical behaviour, generally there are two different methods that have been used to integrate the electrical activity of the tissue into a biomechanical model: (i) as a simulated quantity where the output of an electrical model is input to a separate mechanical model [2], or (ii) as an indirect measure of deformation of the tissue due to electrical contraction [2].

Because the developments of models of electrical and mechanical processes have occurred largely independently, *electro-mechanical coupling*, involving the integration of both techniques, can result in a high computational cost, compromising the efficiency of the model and the scale of the problems that can be addressed. Independent electrical or mechanical models do not easily incorporate the effects of electro-mechanical coupling, such as bending or stretching of ion channels in cellular models, which has been shown to be important in tissue electro-mechanics [1], [3], [5], [6]. Integrating the mechanical response of the tissue into models for electrical activity, and doing so in a computationally efficient way, is, therefore, a key challenge for the generation of models that can advance our fundamental understanding of electro-mechanics in biological tissue [3].

Against this backdrop of limitations, the objective of this work is to present an alternative approach to model the coupled electro-mechanical phenomena of biological tissue in a single model, establishing this approach for the case of a single nervous cell. Electro-thermal equivalence is proposed here for modelling the simultaneous inter-dependence of electrical and mechanical phenomena by using advanced facilities for coupled thermo-mechanical models provided in existing computational modelling tools [7]–[9]. Abaqus CAE 6.13-3 is used in this work. In contrast to previous works using analogue quantities in modelling [7], [8], here, the equivalences are utilised to (i) establish a thermal analogue implementation of the Cable Equation simulating the non-linear diffusion of electrical charges in nervous cells by using the Heat Equation [9]–[11], and (ii) implement coupled electro-mechanical effects by applying a thermal expansion coefficient to describe the piezoelectric nature of the nerve membrane [8], [12].

Analysis of neural signalling is normally carried out with computational software and open source electrical software (such as NEURON and NEURITE [13]) for simulating electrical signal propagation. These software codes solve 1D electrical simulations, in which the material structure of the tissue is approximated or neglected. Hence, advanced electro-mechanical modelling of material characteristics cannot be carried out because of the missing 3D structure. In particular, capabilities for coupled electro-mechanical simulations are not readily available in computational software, but they are well developed for thermo-mechanical analysis. The thermo-electrical analogy therefore enables us to build on these established tools to gain insights into neural activity. The well-established nature of thermo-mechanical analysis ensures that the tools we are using are reliable, and can be reproduced and adopted in similar application fields. In addition, the approach enables modelling of fully coupled 3D electro-mechanical systems.

In contrast to other commercial Multiphysics software, such as COMSOL [14], Abaqus CAE 6.13-3 is an engineering design software that has been developed specifically to account for an extensive range of complex material constitutive relationships in the context of non-linear finite element (FE) analysis. In addition, Abaqus is extensively used in bioengineering thanks to the capabilities it provides for extracting detailed results at different levels in multiscale FE models. Although there is an option in Abaqus for simulating electromechanical systems, this tool is mainly use to model piezoelectric material and it has limited capabilities for defining non-linear electrical properties as required to implement the Hodgkin-Huxley (HH) model [15]. Here, an electro-thermal analogy is proposed to enable more powerful neural modelling, where 3D coupled (equivalent) electro-mechanical models are resolved using established coupled thermo-mechanical computational tools. In, the equivalent electro-mechanical coupling in Abaqus enables investigation of complex coupled phenomena at the nerve membrane layer [16], where there is a close link between electrical and mechanical quantities (as in damage, trauma or diseases [6], [17]).

The first aim of this paper is the assessment of electro-thermal equivalence and the identification of equivalent thermal material properties to model the electrical behaviour of biological tissue. The second aim is to investigate the *coupled electro-mechanical deformation* behaviour of nerve tissue, by combining the electro-thermal equivalences with a coupled thermo-mechanical modelling approach that models the biophysical phenomena at the nerve membrane [18].

First, electro-thermal equivalence is demonstrated for a 3D nerve model, and validated against analytical models for a range of non-symmetric boundary conditions under sub-threshold conditions [11], [14], [15]. The electro-thermal equivalent model is then adapted to account for the non-linear electrical behaviour of the nerve membrane, described by Hodgkin and Huxley (HH) dynamics [15], and validated through comparison with space and voltage clamp experiments [15], [19].

Secondly, an equivalent thermal expansion coefficient is applied to model the piezoelectric properties of the nerve membrane to investigate coupled electro-mechanical phenomena accompanying the action potential [16], [18], [20] as a function of physical length scale. Simulated results of deformation are compared with experimental observations of squid giant axon, crab nerve fibre and garfish olfactory nerve fibre [16], [18].

The overall objective here is to establish, validate and demonstrate the performance of, an effective and powerful platform for 3D electro-mechanical tissue modelling that can form the basis of further studies, for example to integrate the effects of trauma and to account for the resulting variations in conduction of signals [6], and to scale up the approach to the nerve bundle, tissue and organ levels.

## II. MATERIALS AND METHODS

### A. Theory of the Equivalence

The Cable Equation and the Heat Equation are partial differential equations of the same parabolic type [7], [10]. Hence, the mathematical solution of the steady state and transient Heat Equation (1) below, can also be applied to the Cable Equation (2) [7], [10].

In the Heat Equation (1), $T$ is the temperature [$°C$ or $K$], $\rho$ is the density [$kg/m^3$], $c_p$ is the specific heat capacity [$J/(kg\,K)$], $k$ is the thermal conductivity [$W/(mK)$], and $Q$ is the heat source density [$W/m^3$]; correspondly, in the Cable Equation (2) applied to the nerve, $V_m$ is the transmembrane potential [$V$], $S_v$ [$1/m$] denotes the surface-volume ratio, $C_m$ [$F/m^2$] is the nerve membrane capacitance, $\sigma$ [$S/m$] is the electrical conductivity of the membrane, and $I_{ionic}$ [$A/m$] is the ionic current.

$$\rho c_p \ \partial T/\partial t - \nabla \cdot (k\nabla T) + Q = 0 \qquad (1)$$

$$C_m S_v \ \partial V_m/\partial t - \nabla \cdot (\sigma \nabla V_m) + S_v I_{ionic} = 0 \qquad (2)$$

By substituting electrical for thermal quantities, the equivalence relationship becomes evident. Table I lists the quantities of the thermal and electrical systems in analogous form, based on the literature [7], [8], [12]. It is interesting to note that while the equivalence between electrical and thermal conductivities is well known, the equivalence between electrical capacitance per unit area and specific heat capacity is not so obvious, although it is important for incorporating the effect of nerve capacitances in the thermal domain.

Coupling of the electro-mechanical effects of the action potential [15] is achieved through modelling of the nerve membrane as a piezoelectric material [16]. By using the electro-thermal analogy [8], [12], the electric field is equivalent to a thermal load, while the piezoelectric constants are equivalent to the thermal expansion coefficients. Therefore, mechanical effects of the piezoelectric properties of the nerve membrane can be incorporated in an equivalent electro-thermal model by defining an appropriate thermal expansion coefficient. The thermo-elastic strain-stress relation is given in (3), where $\{\varepsilon\}$ is the total strain vector, $[\beta]$ is the compliance matrix, $\{\sigma\}$ is the mechanical stress vector, $\{\alpha\}$ is the thermal expansion coefficients vector and $\Delta T$ the temperature difference [8], [12].

$$\{\varepsilon\} = [\beta]\{\sigma\} + \{\alpha\}\Delta T \qquad (3)$$

Based on this relationship, the piezo-elastic relation is given in (4) where $\{\delta\}$ is the piezoelectric strain coefficient vector, $h$ is the thickness of the piezoelectric layer and $\Delta V$ is the voltage difference [8], [12].

$$\{\varepsilon\} = [\beta]\{\sigma\} + \{\delta\}(\Delta V / h) \quad (4)$$

Simulating the nerve membrane as the dielectric component of a parallel plate capacitor [8], [15], the displacement follows the gradient of the electric field, approximated here as the voltage across the membrane divided by its thickness, using the approach presented in [8], [12], see (4).

TABLE I
ELECTRICAL-THERMAL EQUIVALENCE

| Electric Quantities | Symbol | Units | Thermal Quantities | Symbol | Units |
|---|---|---|---|---|---|
| Voltage | $V$ | [$V$] | Temperature | $T$ | [°C or K] |
| Current per unit area | $I$ | [$A/m^2$] | Heat Flux | $q$ | [$W/m^2$] |
| Charge | $\Phi$ | [$C$] | Energy | $E$ | [$J$] |
| Electrical Conductance per unit area | $g$ | [$S/m$] | Thermal Conductance per unit area | $c$ | [$W/(K\ m^2)$] |
| Electrical Resistance per unit area | $R$ | [$\Omega\ m^2$] | Thermal Resistance per unit area | $R_{th}$ | [$(m^2 K)/W$] |
| Electrical Resistance per unit length | $r$ | [$\Omega/m$] | Thermal Resistance per unit length | $r_{th}$ | [$K/(mW)$] |
| Electrical Capacitance per unit area | $C$ | [$F/m^2$] | Specific Heat Capacity | $c_p$ | [$J/(kgK)$] |
| Electrical Conductivity | $\sigma$ | [$S/m$] | Thermal Conductivity | $k$ | [$W/(mK)$] |
| Temperature Coefficient | $\alpha$ | [$1/V$] | Expansion Coefficient | $\alpha$ | [$1/K$] |
| $C_m S_v$ | $\beta_e$ | [$(C\ V)/m^3$] | $c_p\rho$ | $\beta_t$ | [$(J\ K)/m^3$] |
| Piezoelectric Strain Coefficient | $\delta$ | [$m/V$] | Thermal Expansion Coefficient | $\alpha$ | [$1/K$] |

*B. Neuron Structure*

Neurons or nervous cells are electrically excitable cells carrying electrical signals, called action potentials (or spikes). Each neuron consists of a long cylindrical nerve fibre called the axon, see Fig. 1, which connects between the neuron extremities or dendrites [10]. A thin membrane sheath separates the axon (i.e. axoplasm or intracellular media, ICM) from the external saline solution (extracellular media or extracellular space, ECM). According to previous publications [11], [14], a neurite refers to a part of the body of a nervous cell (axon) that consists of two structurally distinct regions (axoplasm and membrane).

Looking to the literature, the simplest model of a neuron is a 1D uniform cylindrical axon as described by the 1D Cable Equation, (2), where the current flow is predominantly inside the core along the cylindrical axis and the extracellular resistivity is negligible (extracellular iso-potentiality) [10]. Previous studies have been performed on the mathematical derivation of the 3D Cable Equation, where longitudinal and transversal stimuli can be modelled to simulate the exchange of electrical charges in one 3D fibre, when located within a bundle of fibres [11], [14]. Due to the 3D geometry, a finite thicknesses for both the ICM and ECM is introduced, whereas the 1D Cable Equation only considers the axon thickness [10]. In the 3D case, the confined thickness of the ECM represents the physical distance between fibres within a bundle [11], [14], [21], and it is known to have similar finite electrical conductivity as the ICM [11], [14], [22]. Thus, when combining the Cable Equation assumptions with the confined ECM thickness assumption, the extracellular resistance per unit length ($r_{\text{ECM}}$) is much greater than the intracellular resistance ($r_{\text{ICM}}$), i.e. $r_{\text{ECM}} \gg r_{\text{ICM}}$ [11], [21], [23], for better accuracy of the analytical solutions (i.e. with an error of less than 5%) where current is predicted to flow predominantly along the fibre axis [14].

Recent experimental evidence has uncovered complex interactions of biophysical phenomena at the membrane layer in nervous cells during signalling [5], [16]. Since it has been found that the membrane is the main piezoelectric component of the neurite [18], it is imperative to accurately model its structure in coupled electro-mechanical studies. In contrast to [4], [14], in this work, the membrane layer has a finite thickness, so the thin layer approximation for the membrane is not assumed. Hence, here, an isotropic 3D cylindrical model of a nervous cell (including ICM, membrane and thin ECM) is modelled in FE as a cable of semi-infinite length to simulate the charge distribution across an unmyelinated neurite [4], [10], see Fig. 1.

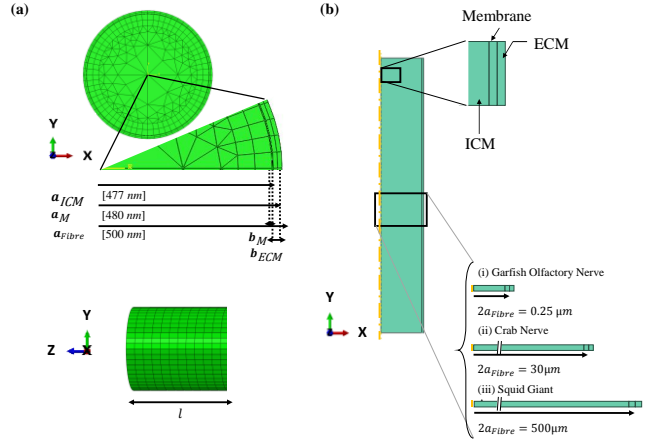

Fig. 1. In (a), a schematic illustration of the unmyelinated nerve and myelinated nerve. ECM, membrane and ICM are highlighted. The length of the fibre is $l$. The radius of a nerve fibre from the ICM to the ECM is $\boldsymbol{a}_{Fibre}$. In (b), frontal and lateral views of the mesh of the 3-layer cylindrical model of the nervous cell with 500 µm length are shown. The ICM and membrane radii are $\boldsymbol{a}_{ICM}$ and $\boldsymbol{a}_M$ respectively. The membrane and ECM thicknesses are $\boldsymbol{b}_M$ and $\boldsymbol{b}_{ECM}$, respectively. In (b), the 3-layer axisymmetric model of the membrane. Below, the 3-layer axisymmetric models of squid giant axon, crab and garfish olfactory nerve fibre with diameters equal to 500 $\mu m$ [10], [15], 30 $\mu m$ [19] and 0.25 $\mu m$ [24] respectively.

The length of the cylindrical model, $l$, see Fig. 1 (a), was chosen to be at least ten times the electronic length constants [14], here 500 $\mu m$, see Fig. 1 (a). Then the radii are: $a_{ICM} = 0.477$ µm for the ICM, $a_M = 0.480$ µm for the membrane, and $a_{ECM} = 0.500$ µm for the ECM [11], [21], see Fig. 1 (a). The diameter of this fibre is within the range of the human optic axon [25]. Here, the membrane's thickness is considered to be that of the central dielectric core of a planar capacitor (about 3 $nm$), without polar head groups [21]. The size of each cross-section is chosen to be within the range of physiological values [14], [23] for a necessary comparison with the analytical solutions [14], [23].

TABLE II
MATERIAL ELECTRICAL PROPERTIES

| Material Properties | Units | ICM | Membrane | ECM |
|---|---|---|---|---|
| Electrical Conductivity | $S/\mu m$ | $10^{-5}$ | $3.14 \cdot 10^{-13}$ | $10^{-5}$ |
| Electrical Capacitance | $F/\mu m^2$ | $1.48 \cdot 10^{-15}$ | $10^{-14}$ | $3.54 \cdot 10^{-14}$ |
| Density | $kg/\mu m^3$ | $10^{-18}$ | $10^{-18}$ | $10^{-18}$ |

*C. Material Properties*

Table II lists the values of the material properties, directly inputted into Abaqus CAE 6.13-3, in which the calculated electrical quantities are assigned to the corresponding thermal equivalent quantities. Due to the constant cross-sectional area, the model is a composite continuous compartment [26].

Both ICM and ECM are purely conductive and homogeneous, see Fig. 1, with equal conductivities [11], [14], [22]. The resistivity of the ICM, $\rho_{ICM}$, and of the ECM, $\rho_{ECM}$, are both set equal to 0.1 Ωm, and derivation of the membrane resistivity is based on a membrane resistance, $R_m$, equal to 10 Ωm² [10], all within the range considered in previous works [11], [23].

The steady state and time varying electrical behaviours are analysed based on the electro-quasi-static (EQS) approximation for diffusion of charge in biological media [7], [27]. The EQS leads to a simplification in Maxwell's equations by ignoring the magnetic field, inductive effects [28], (electro-magnetic) wave propagations at low frequency electro-dynamics (less than 1 $kHz$ [28], [29]) and non-ideal capacitive effects, such as dielectric loss factor [30], [31].

For simplicity, the frequency dependence of the neuron constituent material properties, such as permittivity ($\varepsilon$) and conductivity ($\sigma$), is neglected due to the EQS assumption [26], [32]. This also aligns with previous work where the inclusion of nonhomogeneous media (such as the proposed nervous cell structure) has been shown to accurately account for frequency-dependent behaviour when the transmission of electric signals (such as the action potential) occurs at frequencies $\omega$ higher than 100 Hz ($\omega \varepsilon/\sigma \gg 1$) [26], [28], [30]. The ICM and ECM are assigned relative permittivity values equal to the static relative permittivity of water, i.e. 80 [4], [21]; the value for the membrane is 3.5, based on a capacitance $C_M$ [15] equal to $10^{-14}$ $F/\mu m^2$, see Table II. With constant values for $\sigma$ and $\varepsilon$, the membrane of the neurite is modelled as a distributed (passive) parallel resistor–capacitor structure in the subthreshold regime [11], [21].

The trans-membrane electrical behaviour of a fibre has two distinct conditions of interest: (i) the subthreshold excitation condition (prior to action potential initiation), where the membrane is passive, and (ii) the trans-threshold (near-, upper-threshold or active state) condition, where the membrane is active [21]. In upper-threshold simulations, the membrane conductivity $\sigma_M$ changes in response to the membrane voltage as described by the HH dynamics [15]. Similarly, the membrane electrical capacitance per unit area (and thus the relative permittivity) is a constant [15] for a non-coupled electro-thermal analysis, while it varies to account for electrostriction in the coupled electro-mechanical model [33].

Additionally, in the coupled electro-mechanical model, the piezoelectric effect is only relevant in the through-thickness direction, and we thus assume only one non-zero component for $\delta$, approximately 1 $nm$ per 100 $mV$ [18] in that direction, see (4). As the electrical depolarization is travelling along the nervous cell [15], compressive forces on the membrane arise from the change in shape (electrostriction) [18], [20] and there is a corresponding change in capacitance [33]. In our model, the total charge displaced arising from the electrostriction is proportional to the square of the membrane potential, $V_m^2$ [33].

*D. Subthreshold Model: Mesh and Boundary Conditions*

To ensure numerical accuracy, the minimum FE size is chosen to be smaller than the thickness of the membrane, here 3 $nm$. As shown in Fig. 1 (a), the mesh density changes through the thickness of the intracellular media: the inner core was partitioned with a coarse mesh of the wedge element type, and from a radius of 350 $nm$ outwards, the mesh is more regular and dense with hexahedral elements. The mesh design was motivated by computational efficiency. The membrane and the extracellular media were defined as solid bi-layers within the mesh. In this way, the same hexahedral element was assigned to the regions where higher levels of charge are exchanged. The model consists of 1,276,922 nodes, 1,426,524 elements, where 1,379,904 are linear hexahedral elements (DC3D6 in Abaqus) and 46,620 are linear wedge elements (DCC3D8D) for convection/diffusion analysis with dispersion control.

The practical case of a standard stimulating electrode on a neurite has been previously modelled using electrical FE simulation [11], [14], [23], where the current spreads both along the fibre and across its layers. In order to reproduce this phenomenon, the (passive) membrane is modelled as a homogeneous conducting cylinder with a non-uniform voltage $V$ [$V$], (5), or current density $I$ [$A/m^2$] boundary condition, (6), applied as external stimulation [11], [14]. In (5)-(7), $n$ is the mode number of the input signal, which accounts for variable sinusoidal voltage or current distributions acting along the azimuthal coordinate $\theta$ of the neurite. The temporal frequency $\omega$ acts over time, $t$, as shown in (5) and (6), respectively. The longitudinal width of the current and voltage stimuli is controlled through the standard deviation, $s$, of a Gaussian distribution of the input function along the axial dimension, $z$, according to (7).

$$V_{input}(z,\theta,t,n) = V(n) G(z) \cos(n\theta) \cos(\omega t) \quad (5)$$

$$I_{input}(z,\theta,t,n) = I(n) G(z) \cos(n\theta) \cos(\omega t) \quad (6)$$

$$G(z) = (2\pi)^{-\frac{1}{2}} e^{-z^2/2s^2} \quad (7)$$

At this stage, the magnitude of the voltage and current stimuli is constrained to be within the subthreshold regime, i.e. the trans-membrane voltage contains no action potential.

*E. Active Membrane Model: Mesh and Boundary Conditions*

In order to validate against the HH model [15], the non-linear dynamic behaviour of the membrane is simulated assuming axisymmetric uniformity of the potential (i.e. longitudinal mode); hence, an isotropic axisymmetric model is used, based on a 2D longitudinal section of the full nerve fibre, see Fig. 1 (b) [21]. Therefore, in this case all the field quantities are assumed to be independent of the azimuth angle of the cylinder. Here we used 15,652 nodes and 15,300 axisymmetric quadrilateral elements designed for convection/diffusion analysis with dispersion control (type DCCAX4D in Abaqus). Thanks to the symmetry of the geometry and boundary conditions, the use of an axisymmetric geometry reduces the computational cost of the simulation. However, our approach to modelling the non-linear dynamic behaviour is valid for 3D models as well, and as such the 3D model of the nervous cell is also used here, see Fig.1, for the subthreshold model.

The left side of Fig. 2 shows the flowchart for implementing the HH model in which the Gaussian voltage distribution input elicits the action potential, generating a flow of ionic currents across the membrane as described in [15]. On the right side of Fig. 2, the thermal equivalent implementation of the HH model is illustrated for 3D distributions of voltages and currents, which are simulated as temperature and heat flow distributions, respectively in the 3D FE thermal model of the nervous cell. The inclusion of the HH dynamics [15], see (8), leads to the non-linear cable equation, where the speed and shape of the solution are defined by the standard HH voltage gating variables ($\bar{g}_K, \bar{g}_{Na}, \bar{g}_l; V_K, V_{Na}, V_l$ —see Sections A and C in the supplementary material for further details).

$$\frac{a_{ICM}}{2R_{ICM}v^2}\frac{\partial^2 V}{\partial t^2} = C_M \frac{\partial V}{\partial t} + \bar{g}_K n^4 (V - V_K) \qquad (8)$$
$$+ \bar{g}_{Na} m^3 h (V - V_{Na})$$
$$+ \bar{g}_l (V - V_l)$$

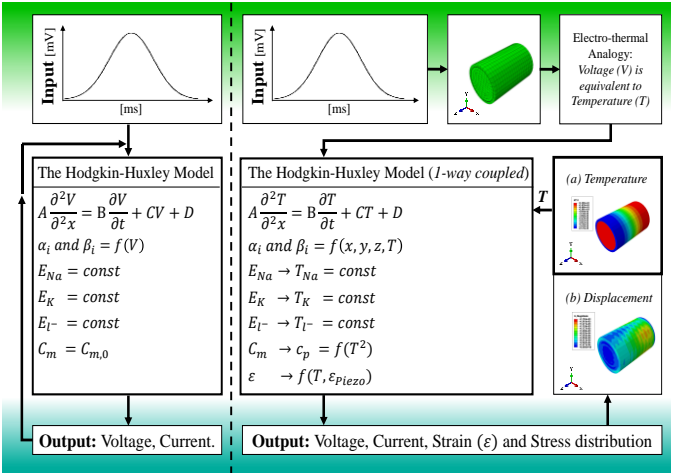

Fig. 2. Flowchart of the code describing the active behaviour of the nerve's membrane: on the left the HH dynamics [15] and on the right the 1-way coupled HH dynamics. Here, a Gaussian voltage distribution elicits the action potential in a 3D model of a nervous cell. By using electro-thermal equivalences, the HH dynamics are implemented as an equivalent thermal process, in which the membrane's conductivity changes as in [15] and the capacitance, $C_m$, changes as in [33].

In Abaqus CAE, the non-linear behaviour of the membrane is implemented using user-defined subroutines to describe the changes in membrane thermal equivalent conductivity with membrane voltage, as in [15] (UMATHT, USDFIELD and DISP subroutines in Abaqus).

The linear Cable Equation (2) and the Heat Equation (1) are diffusive equations with no (linear) travelling-wave component, whose solutions correspond to passive, diffusive spreading [34]. Here, in the equivalent electro-thermal domain, the action potential is a localized voltage distribution along the fibre in a dissipative system [35], described by an RC circuit, according to the EQS assumption [26]. If required, the propagation of the action potential can be implemented as a moving heat source at the membrane, derived from a wave-like solution of the Heat Equation.

*F. Equivalent Coupled Electro-Mechanical Model: Mesh and Boundary Conditions*

Numerous mechanical events have been experimentally observed in neuronal membrane excitability and are thought to play an important role in regulating neuronal activity [5], [16], [20]. Herein, we assess the electro-thermal equivalence as a method for investigating the mechanical displacement arising from neural transmission. In this model, the mechanical features are electrically driven phenomena, implemented through a thermal expansion coefficient that represents the piezoelectric effect [20] as illustrated in Fig. 2.

The equivalent electro-mechanical coupling is assessed for three different models, simulating the electro-mechanical responses of isolated unmyelinated fibre of the squid giant axon, crab nerve fibre, and garfish olfactory nerve fibre with diameters equal to $500\ \mu m$ [10], [15], $30\ \mu m$ [19] and $0.25\ \mu m$ [24] respectively, see Fig. 2. As in [20], the width and amplitude of the Gaussian distribution function of the excitation voltage for each fibre is dependent on fibre dimensions, as shown in Fig. 5.

Homogeneous, isotropic and incompressible mechanical properties are assigned to the ICM, membrane and ECM for each nervous cell type [20]. The density of the medium is set to $10^3 kg\ m^{-3}$, i.e. the same as water [20]. Size-independent elastic material properties are assigned to each animal. In Case I, each nervous cell is assumed to be made of different regions with the same mechanical properties [20], [33] (where Young's Modulus [33] is $1.4\ \times 10^8 Pa$). In Case II, only the membrane and the ICM have the same Young Modulus ($63,981.69\ Pa$ [36]), while the ECM has a different value ($23,195.14\ Pa$ [36]). This allows for the effects of elastic properties to be assessed.

Each nervous cell type is modelled with an axisymmetric geometry assumption, simulating a nerve fibre with semi-infinite length. The minimum FE size is $2.14\ \times 10^{-4}\ \mu m$ at the membrane layer. The models consist of: 124,031 nodes and 120,000 4-node axisymmetrical coupled thermo-mechanical elements (CAXA4RT) for the garfish olfactory nerve; 33,033 nodes and 32,000 CAXA4RT elements for the crab nerve fibre and 17,535 nodes and 170,000, CAXA4RT elements for the squid giant axon. An encastre boundary condition is enforced at the origin of each model, while the longitudinal displacement is limited along the axis of symmetry and for the nodes at $y = 0$, see Fig. 2.

## III. Results

Electro-thermal equivalences are validated for a 3D neuron model (see Fig. 1 (a)) in both subthreshold and upper-threshold regimes with non-symmetrical voltage and current boundary conditions. Then, the electro-mechanical coupling is validated under upper-threshold conditions.

In the reported results, the trans-membrane potential is referred to as the membrane potential for subthreshold stimulation and the action potential for upper-threshold stimulation. Through the thermo-mechanical analogy, the temperature (NT11) representing voltage is a nodal quantity, determined though the TEMP variable in Abaqus CAE.

### A. Subthreshold Model

Fig. 3 shows the 3D model results for the steady state voltage boundary conditions of the nerve membrane. Here, only the results for the subthreshold model with $n = 2$ are shown. In Fig. 3 (a), the contour plots show a part of the neurite's section with $5\ \mu m$ and $15\ \mu m$ length, respectively. In Fig. 3 (a) and (b), a uniform voltage boundary condition (5), $(V(2) = 20\ nVm)$ with a Gaussian distribution (7), was applied at the outer surface of the ECM ($s = 1\ \mu m$), similar to that applied in [14]. In Fig. 3 (b), the voltage distribution in the cylinder in the axial direction is shown, and compared with results produced by the analytic solution presented in [9], where close agreement between both models is confirmed. Similarly, results for current boundary conditions can be found in the supplementary material.

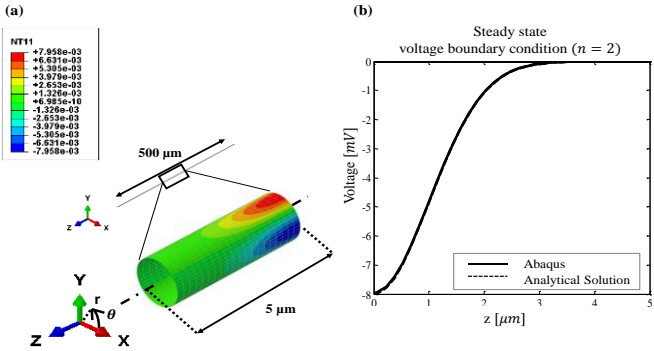

Fig. 3. Voltage boundary condition. (a) Spatial variation of the membrane voltage [V], here NT11, in steady state ($\omega = 0$) is shown for $n = 2$ [14], see (5). The azimuthal coordinate is $\theta$. In (b), the voltage ditribution in the axial direction at the membrane layer, at $x = 1\ \mu m, y = 0\ \mu m$, from $z = 0\ \mu m$ to $z = 5\ \mu m$ of the 3 layer cylinder model. Here $V(2) = 20\ nVm$ and $s = 1\ \mu m$ [14], see (5) and (7).

### B. Active Membrane Model

The equivalent implementation of the HH dynamics has been validated according to the voltage clamp and space clamp procedures [15], applying a clamped voltage, Fig. 4 (a), and current stimulus, $I$, Fig. 4 (c), respectively. The sodium, $I_{Na}$, potassium, $I_K$, and leak, $I_l$, currents are the ionic currents of the nerve membrane, see Fig. 4 (d), as in [15]. The generated ionic membrane currents are within physiological values [15] and are reported in Fig 4 (b) and (d), respectively, for each input. Additional results about the gating variables can be found in the supplementary material.

### C. Equivalent Electro-Mechanical Coupling

The action potential is, here, implemented as a voltage source, whose amplitude is a Gaussian distribution with fixed spatial width that depends on the nerve fibre dimensions, as in [15], [20], see Fig. 5.

For each of the three test cases considered, we apply the distribution of the action potential as an equivalent thermal excitation to calculate the membrane displacements, stresses and strains in the middle of the membrane layer (at the peak of the action potential), see Fig. 5. A decrease in thickness of about 23% is found at the membrane, consistent with what is experimentally observed (16% [16], [37] and 21.6% [38]). Moreover, the change in the capacitance is proportional to the voltage square changes across the membrane (calculated approximately as 0.03% during the action potential), once again consistent with physiological conditions (less than 1% [33]). Although changes in membrane density strongly affect the membrane capacitance during signalling [16], [37], here, the capacitance per unit area is a function of the voltage square only, as in [20], [33], where the initial estimate of the membrane capacitance is $0.75\ \mu F/cm^2$ [33]. Hence, the membrane deformation is dominated by the piezoelectric effects, with a maximum deformation magnitude of approximately 23% of the membrane thickness.

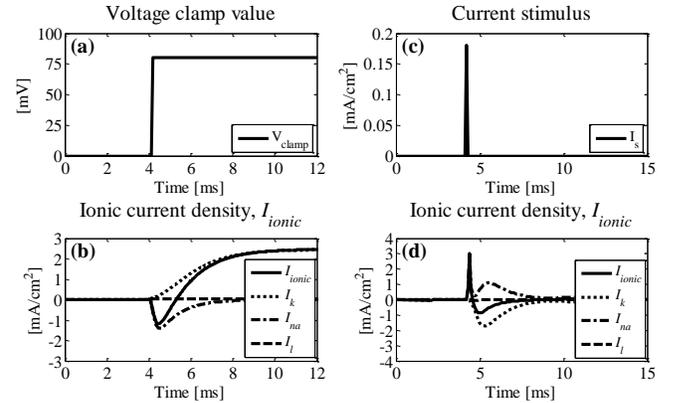

Fig. 4. Voltage clamp procedure with $80\ mV$ clamped voltage, (a), and generated ionic currents in (b). Space clamp procedure and $18\ mA\ cm^{-2}$ current stimulus of $0.2\ ms$ (c) and generated ionic currents in (d).

In Fig. 5 (d), the time-dependency of the action potential refers to the squid giant axon only, which is in agreement with experimental evidence (from approximately $-65\ mV$ to $+40\ mV$ [15], [20]). Similar changes and values are found both for the garfish and crab nerve fibres. In contrast to [20], [36], as a first step to assess the performance of a fully coupled electro-mechanical model, incompressible isotropic mechanical properties are used to characterize each axon model while the piezoelectric effect is only relevant in the through-thickness direction of the membrane layer. Hence, the electrically driven displacement and strain each have a predominant radial component in the membrane. The assumptions focus the coupled electro-mechanical analysis on the assessment of the theory of equivalences and the impact of equivalent thermal phenomena on the structure of a single nervous cell.

Fig. 5 (a)-(c) shows results for a Young Modulus of $1.4 \times 10^8\ Pa$ (Case I) [33], and Fig. 5 (d) compares the stresses and strains estimated using both Case I [33] and Case II [36] over time. It is interesting to note in Fig. 5 (d) that while the strains, which are electrically driven, are quite independent of Young's Modulus, the opposite is true for the stresses, which are highly dependent on Young's Modulus.

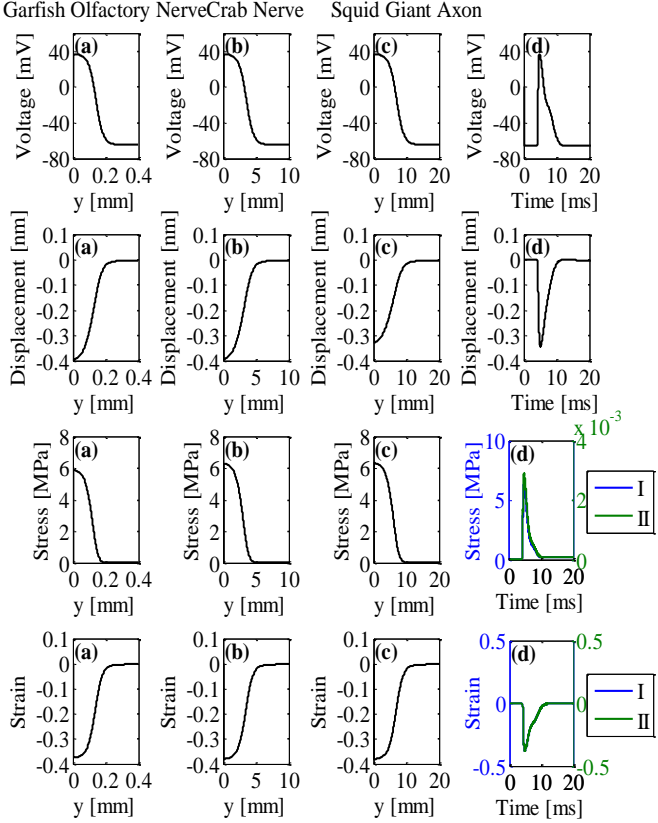

Fig. 5. Quantitative comparison with three experimentally favored systems of garfish olfactory nerve (column a), crab nerve (column b) and squid giant axon (column c) using model and parameters as described in the text. Columns (a)-(c) present the spatial distribution of the voltage and mechanical responses for the three different cases. Column (d) presents the time dependency of the mechanical features for the squid giant axon case only. The results refer to the 1-way coupling of the HH model. The results are taken in the middle of the membrane layer at $y = 0$, where the $y$−axis is the axis of symmetry of the model and the $x$−axis is the radial distance. For each case, the width of the action potential is taken from experimental literature [20]. Displacement, strain and stress are taken in the radial direction. In (d), stresses and strains are estimated using different Young's Moduli, refer to Case I [20], [33] and Case II [36].

## IV. DISCUSSION

The main achievements of this paper are: the analogue implementation of the Heat Equation as the Cable Equation; the validation of electro-thermal equivalences in an uncoupled electrical and a coupled electro-mechanical analysis; and the modelling of a nervous cell as a 3D 3-layer conductor at different physical scales. Previous works have introduced only the use of analogue quantities in finite element modelling [7], while others [11], [14], [23] neglect the importance of the electro-mechanical biophysical phenomena at the membrane layer (as electrostriction, swelling and deformation) [5], [16], [38]. The assessment and validation of the electro-thermal equivalences and equivalent material properties in Abaqus CAE 6.13-3 are necessary to guarantee that the electrical current flow in a 3D neuron model can be represented as a heat diffusion process under longitudinal and transversal stimuli, for both passive and active membranes. Here, the Heat Equation is used to simulate not just analogue quantities [7] but also an equivalent non-linear phenomenon, specifically addressed in the Core Conductor Theory [10] and represented by the Cable Equation for nerve fibres [10]. Validation of the solution for the Cable Equation in turn enables development of a coupled electro-mechanical model through the development of an equivalent coupled thermo-mechanical model.

When compared with other approaches for coupled electro-mechanical modelling, the combined real-time use of separate advanced mechanical and electrical modelling software codes is computationally very expensive and computationally intractable if inhomogeneity and anisotropy in tissue conductivities are considered. Here the analogue thermo-mechanical domain provides computational simplification, particularly for three-dimensional problems in FE.

The simulation of electrical conduction using the electro-thermal equivalence can be implemented in different FE software packages (e.g. ANSYS and COMSOL), and non-linear electrical activity can be implemented directly in COMSOL [4], but coupled electro-mechanical solutions to biological systems have not yet been reported for this software. However, the use of Abaqus in the current application presents significant advantages. Compared to COMSOL, Abaqus provides greater control over the range of finite element types used in the model, the mesh structure, the calculated field variables, and the extraction of detailed field variable values at different levels in multi-scale models. The use of Abaqus also facilitates the use of a very broad range of complex non-linear electrical and mechanical material properties, including user-defined material descriptions, which lend themselves particularly well to the representation of biological tissue. Hence, in contrast to previous studies [4], [14], [27], [39], the advanced FE platform for material characterization makes Abaqus CAE an ideal tool for investigating the electro-mechanical coupling in nerves.

The non-linear Cable Equation considered in this paper focused on the non-linear HH dynamics for an unmyelinated nervous cell [15]. However, the same equation can be used for a myelinated nerve fibre where the conductivity is periodic and piecewise constant, as is the case for Ranvier's nodes for example [40].

In relation to the application of electro-thermal equivalences to solve coupled electro-mechanical nerve conduction problems, very good agreement between the simulated results and published experimental measurements (over a number of quantities) validates the approach taken, and in particular the assumption of approximating the voltage gradient (equation (4)) [8], [12]. Secondly, by using three different models (squid giant axon, crab nerve fibre and garfish olfactory nerve fibre), the results confirm the validity of the approach over three very different length scales. It is interesting to note that although the electrical properties are different in each case, the strain response exhibits a similar trend across the different scales. This can be understood when one considers that the piezoelectric component of the strain is directly related to the piezoelectric constant (assumed the same for all cases) and the action potential which is relatively uniform across all the cases (see Fig. 5 (a)). Finally, although strain levels are similar across the scales, the stress levels are hugely dependent on the choice of Young's Modulus. This demonstrates the importance of having accurate tissue mechanical property data in such models for gaining accurate insights into the coupled electro-mechanical behaviour. Additionally, dependence of electrical phenomena on stress through two-way coupling is expected to be significant.

Our model assumes that the mechanical features of the membrane are electrically driven by the action potential [15], [20]. The membrane is modelled as a parallel plate capacitor [8], [15], [16], in which the piezoelectricity follows the orientation of the electric field [18] and the changes in capacitance are proportional to the square of electric polarization [33]. Although the HH model [15] assumes a constant membrane capacity, the combined use of the depolarization wave and the electrostriction seems to be accepted in literature [20]. Isotropic material properties are assumed in all regions (apart from the orthotropic thermal expansion coefficient in the membrane layer). Analysis with a set of non-isotropic mechanical properties at the membrane layer [16], [37] would require the use of other user-defined subroutines in Abaqus.

In conclusion, a method for modelling the mechanical characterization of neural activity in a coupled electro-mechanical domain by using the electro-thermal equivalences in FE has been proposed and validated. In contrast with [4], [14], [20], [27], [39], this approach can be used to better investigate the combined impact of the electric field and mechanical stresses simultaneously in the alteration of neural patterns and diseases [41]. Thus, we stress the importance of electro-thermal equivalences through the analysis of the electric current in a nervous cell, concluding that the physical meaning of the electric conduction is not altered using an analogue domain. This coupling is based on a 1-way coupled HH model in which material properties vary only in response to the simulated voltage (equivalent to temperature) during the cycle. Work is on-going to extend the approach to modelling of biological systems at the tissue and organ levels and to model the neural activity in injured fibres and bundles [6], where the material properties change in response to voltage and strain.

V. CONCLUSION

This paper is focused on the assessment and validation of the use of electro-thermal and electro-mechanical equivalences for modelling the inter-dependence of electrical and mechanical phenomena in FE analysis, considering the case of a single nervous cell. This approach facilitates the modelling of electro-mechanical phenomena, opening the way for realistic 3D electro-mechanical models of neural networks, neuropathological, neurological traumatic brain injuries and diseases (such as multiple sclerosis) [6], [17].

The model is validated against subthreshold and upper-threshold stimuli and is capable of simulating the electric activity of a nervous cell with both voltage and current boundary condition in the steady and transient state. The electrical conduction is well described by this approach, which verifies the spatio-temporal variation of the electro-thermal equivalences by using the Heat Equation as an analogue of the Cable Equation. In the end, the electro-mechanical coupling arises from the mechanical features of the Cable Equation which, here, is used for simulating both piezoelectricity and electrostriction.

In the context of multi-disciplinary modelling, this methodology can help the integration and interaction of the mechanical modelling of materials with electrical activity for general applications. The inconvenience of having to combine different software packages and codes can therefore be overcome, by utilising the thermal analogue of electrical behaviour in the same modelling software, i.e. Abaqus CAE 6.13-3, that releases full access to both mechanical and equivalent electro-thermal quantities at each node and element of the model.

ACKNOWLEDGMENT

The authors acknowledge funding from the Galway University Foundation, the Biomechanics Research Centre and the Power Electronics Research Centre, College of Engineering and Informatics, National University of Ireland Galway.